\documentclass[aps, prx, nofootinbib, twocolumn,amsfonts,amsmath,amssymb,mathtools,superscriptaddress,preprintnumbers,floatfix,reprint]{revtex4-2}

\usepackage{graphicx}
\usepackage{subcaption}
\usepackage{microtype}
\usepackage[hidelinks,colorlinks=true,linkcolor=blue,citecolor=blue]{hyperref}
\usepackage[utf8]{inputenc}
\allowdisplaybreaks
\usepackage{mathtools}
\mathtoolsset{centercolon}
\usepackage{ragged2e}
\usepackage{bbm}
\usepackage{tikz}
\usetikzlibrary{arrows.meta}

\newcommand{\alg}{\mathfrak{g}}

\DeclareMathOperator{\sn}{sn}
\DeclareMathOperator{\cn}{cn}
\DeclareMathOperator{\dn}{dn}
\DeclareMathOperator{\rank}{rank}
\DeclareMathOperator{\tr}{Tr}

\tikzset{
  sol/.style     = {circle, draw=black!55, fill=black!12,
                    minimum size=5.5pt, inner sep=0pt},
  Sedge/.style   = {blue!70!black, thick, <->, >={Stealth[length=4pt]},
                    shorten <=2pt, shorten >=2pt},
  Tedge/.style   = {red!75!black,  thick, <->, >={Stealth[length=4pt]},
                    shorten <=2pt, shorten >=2pt},
  Sloop/.style   = {blue!70!black, thick, ->, >={Stealth[length=4pt]},
                    shorten <=1pt, shorten >=1pt},
  Tloop/.style   = {red!75!black,  thick, ->, >={Stealth[length=4pt]},
                    shorten <=1pt, shorten >=1pt},
  Tdir/.style    = {red!75!black,  thick, ->, >={Stealth[length=4pt]},
                    shorten <=2pt, shorten >=2pt},
  sollbl/.style  = {font=\small,      inner sep=2pt},
  edgelbl/.style = {font=\scriptsize, inner sep=1.5pt},
}

\makeatletter
\newcommand{\otherlabel}[2]{\protected@edef\@currentlabel{#2}\label{#1}}
\makeatother

\begin{document}

\title{Black Holes, the Bethe Ansatz, and Elliptic Calogero--Moser Systems}
\author{Marco Fazzi}
\author{Kuba Krawczyk}
\affiliation{School of Mathematical and Physical Sciences,
University of Sheffield, Sheffield S3 7RH, UK}

\begin{abstract}
We define a map from solutions of the Bethe Ansatz equations (BAEs) of
four-dimensional $\mathcal{N}=4$ super-Yang--Mills with arbitrary semisimple
gauge algebra $\alg$ to extrema and poles of the potential of the untwisted
elliptic Calogero--Moser system of type $\alg$. We conjecture the map to be a
bijection on the preimage of the Calogero--Moser extrema, and show that it
intertwines the symmetries of the two systems, both the gauge ones (torus, Weyl
and center invariance) and a $\mathrm{PSL}(2,\mathbb{Z})$ action, so that
solutions on both sides organize into orbits, each BAE orbit mapping onto a
single orbit of Calogero--Moser extrema or poles. That the system produced is
the untwisted one has a consequence: the conjectured correspondence between BAE
solutions and vacua of the $\mathcal{N}=1^\ast$ deformation of $\mathcal{N}=4$, which are extrema of the \emph{twisted} system, cannot
extend to non-simply-laced $\alg$. It also fails within the simply-laced cases, though not for $\mathfrak{su}(N)$: we exhibit a few $\mathfrak{so}(8)$ solutions that flow to poles of
the Calogero--Moser potential rather than to extrema, and so have no
$\mathcal{N}=1^\ast$ counterpart. We illustrate the map in detail for every
rank-two $\alg$, classical and exceptional alike, and use these cases as
evidence for the conjecture.

\end{abstract}

\maketitle

%%%%%%%%%%%%%%%%%%%%%%%%%%%%%%%%%%%%%
\section{Introduction and motivation}
\label{sec:intro}
%%%%%%%%%%%%%%%%%%%%%%%%%%%%%%%%%%%%%

The correspondence between supersymmetric gauge theories and classical
integrable systems is one of the oldest threads in mathematical physics.
It begins with Seiberg--Witten theory \cite{Seiberg:1994rs,Seiberg:1994aj},
whose low-energy dynamics was recognized to be governed by an algebraically
integrable system \cite{Gorsky:1995zq,Martinec:1995by,Donagi:1995cf}: the
Seiberg--Witten curve and differential are the spectral data of that system,
which for four-dimensional (4D) $\mathcal{N}=2^\ast$ super-Yang--Mills (SYM) is the elliptic
Calogero--Moser (CM) system
\cite{Calogero:1975ii,Moser:1975qp,Olshanetsky:1981dk,Donagi:1995cf}. The
dictionary is sensitive to the gauge algebra $\alg$ in a way that will matter to us
below. For non-simply-laced algebras the relevant system is not the CM system
built naively on the root system of $\alg$, but its \emph{twisted} counterpart,
associated with a twisted affine algebra
\cite{Inozemtsev:1989,DHoker:1998zuv,Bordner:1998sw,Bordner:1998xs}. (That the twisted systems are 
the ones governing $\mathcal{N}=2^\ast$ with a general gauge algebra was
established through the Lax-pair and spectral-curve construction of
\cite{DHoker:1998yga,DHoker:1999sqz,DHoker:1999hmo,DHoker:1998zuv,DHoker:1998rfc,DHoker:1998xad,DHoker:1997hut}; see e.g. \cite{DHoker:1999yni}
for a review.) The prototype is Inozemtsev's $BC_N$ system, in which the couplings sit at
all four half-periods of the torus \cite{Inozemtsev:1989}. The class of
integrable systems realized in this way is in fact wider than the root systems
of simple Lie algebras: the Inozemtsev system is the Seiberg--Witten system of
4D $\mathcal{N}=2$ $\mathfrak{usp}(2N)$ with four fundamentals and an
antisymmetric hypermultiplet \cite{Argyres:2021iws}, an instance of a general
correspondence between crystallographic elliptic Calogero--Moser systems,
classified by complex crystallographic reflection groups rather than by root
systems alone, and Seiberg--Witten integrable systems \cite{Argyres:2023tfx}.

A second thread is the Bethe/gauge correspondence
\cite{Nekrasov:2009ui,Nekrasov:2009uh}, which identifies the supersymmetric
vacua of a gauge theory with the Bethe states of a quantum integrable system,
the vacuum equations becoming Bethe Ansatz equations. A third thread, closer to the observables we study, runs through supersymmetric partition functions themselves: the superconformal index of class-$\mathcal{S}$
theories computes eigenfunctions of the elliptic Ruijsenaars--Schneider model
\cite{Gaiotto:2012xa,Razamat:2013jxa,Rastelli:2014jja}, a relativistic
deformation of Calogero--Moser, with the residues of the index at flavor poles
extracted by difference operators \cite{Gaiotto:2012xa}. (That line remains active \cite{Nazzal:2023bzu,Kim:2024hxo,Deb:2025ypl,Zeev:2026alb}.)

The threads meet in the study of the index itself. Building on the 4D A-model of
\cite{Closset:2017bse}, Benini and Milan \cite{Benini:2018mlo} recast the index
of a 4D $\mathcal{N}=1$ theory as a finite sum over the solutions of a set of
transcendental equations, dubbed Bethe Ansatz equations (BAEs), each solution
weighted by the integrand and the inverse Jacobian of the equations themselves.
This reformulation proved decisive for the supersymmetric anti-de Sitter (AdS)
black-hole microstate counting problem: evaluating the sum at large $N$ for 4D
$\mathcal{N}=4$ $\mathfrak{su}(N)$ SYM reproduces the Bekenstein--Hawking
entropy of the rotating, electrically charged BPS black holes
\cite{Gutowski:2004ez,Gutowski:2004yv,Chong:2005da,Chong:2005hr,Kunduri:2006ek}
embedded in the dual AdS$_5\times S^5$ vacuum, thus resolving a puzzle that had
stood for over a decade \cite{Benini:2018ywd}.\footnote{The Bethe Ansatz route
is not the only one. The entropy is equally recovered from a Cardy-like limit of
the matrix-model representation of the index (see \cite{Choi:2018hmj,Cabo-Bizet:2018ehj,Cabo-Bizet:2019osg,Cabo-Bizet:2020nkr,Honda:2019cio,ArabiArdehali:2019tdm,Cabo-Bizet:2019osg,Kim:2019yrz,Amariti:2020jyx,Amariti:2021ubd,GonzalezLezcano:2020yeb,Amariti:2019mgp,Amariti:2024bsr,Amariti:2023rci,David:2020ems} for a partial list of results),
from a direct saddle-point analysis in which the black hole is the $(1,0)$
member of a family of complex saddles labeled by coprime pairs
\cite{Cabo-Bizet:2019eaf,BenettiGenolini:2023rkq}, and from the giant-graviton expansion
\cite{Imamura:2021ytr,Gaiotto:2021xce,Murthy:2022ien,Beccaria:2023hip,Lee:2022vig,Arai:2019aou,Arai:2019wgv,Arai:2019xmp,Arai:2020uwd,GonzalezLezcano:2026wje,Deddo:2025lfm}.
Generalized Cardy limits at rational values of the angular potentials give
instead the entropy rescaled by $1/m$ \cite{Goldstein:2020yvj,Jejjala:2021hlt}.} The index had been computed at large $N$
already in \cite{Kinney:2005ej}, where it was found not to grow rapidly enough
to account for that entropy, the required growth being obscured at real
fugacities by boson--fermion cancellations and becoming manifest only once they
are taken complex. Solving the BAEs is therefore of direct physical interest,
and yet explicit solutions have been obtained almost exclusively for gauge
algebras of type $A$
\cite{Hong:2018viz,Benini:2021ano,GonzalezLezcano:2019nca,GonzalezLezcano:2021nzk,ArabiArdehali:2019orz,Lanir:2019abx,Amariti:2025vjd,Hosseini:2016cyf}.
Solutions beyond type $A$ were found only recently by the two authors
\cite{B2paper}, for every rank-two semisimple algebra.

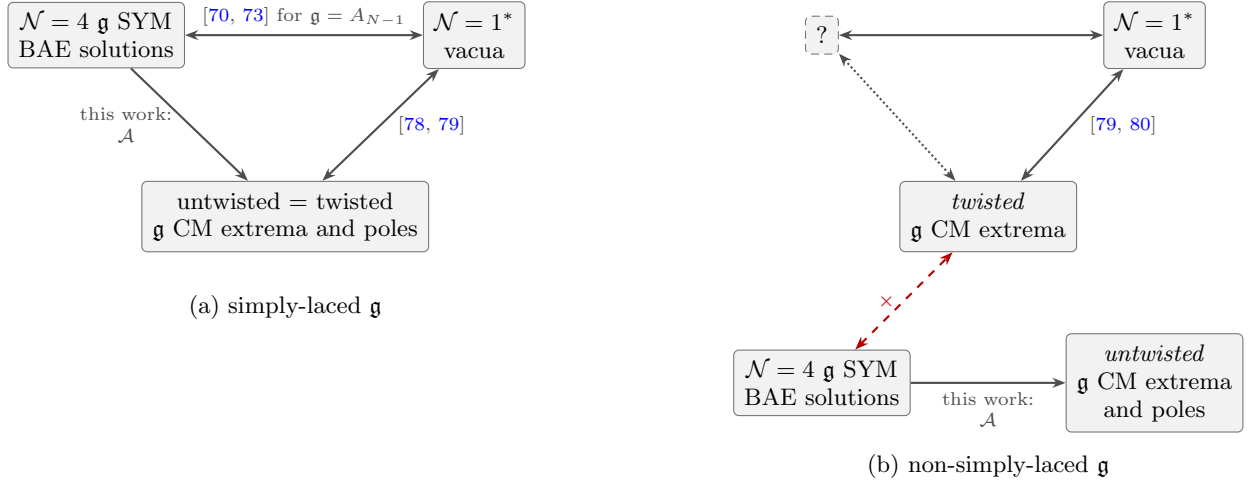
\begin{figure*}[ht!]
\centering
\begin{tikzpicture}[
  vtx/.style   = {draw=black!55, rounded corners=2pt, fill=black!5,
                  align=center, inner sep=4pt, font=\small},
  unk/.style   = {draw=black!55, rounded corners=2pt, fill=black!5,
                  align=center, inner sep=4pt, font=\small, densely dashed},
  leg/.style   = {<->, >={Stealth[length=5pt]}, thick, black!70},
  ours/.style  = {->,  >={Stealth[length=5pt]}, thick, black!70},
  spec/.style  = {<->, >={Stealth[length=5pt]}, thick, black!70, densely dotted},
  cut/.style   = {<->, >={Stealth[length=5pt]}, thick, red!70!black, dashed},
  lbl/.style   = {font=\scriptsize, align=center},
]

%%%% (a) simply-laced
\begin{scope}
  \node[vtx] (bae) at (0,0)     {$\mathcal{N}=4$ $\alg$ SYM\\ BAE solutions};
  \node[vtx] (n1)  at (5,0)     {$\mathcal{N}=1^\ast$\\ vacua};
  \node[vtx] (cm)  at (2.5,-2.4){untwisted $=$ twisted\\ $\alg$ CM extrema and poles};

  \draw[leg]  (bae) -- node[lbl, above]
              {\cite{ArabiArdehali:2019orz,Benini:2021ano} for $\alg=A_{N-1}$} (n1);
  \draw[leg]  (n1)  -- node[lbl, right=3pt]
              {\cite{Dorey:1999sj,Kumar:2001iu}} (cm);
  \draw[ours] (bae) -- node[lbl, left=3pt]{this work: \\$\mathcal{A}$} (cm);

  \node[font=\small] at (2.5,-3.6) {(a) simply-laced $\alg$};
\end{scope}

\begin{scope}[xshift=9.6cm]
  \node[unk] (q)    at (0,0)     {?};
  \node[vtx] (n12)  at (4.4,0)   {$\mathcal{N}=1^\ast$\\ vacua};
  \node[vtx] (cmt)  at (2.2,-2.4){\emph{twisted}\\ $\alg$ CM extrema};
  \node[vtx] (bae2) at (0,-4.6)  {$\mathcal{N}=4$ $\alg$ SYM\\ BAE solutions};
  \node[vtx] (cmu)  at (4.4,-4.6){\emph{untwisted}\\ $\alg$ CM extrema \\ and poles};

  \draw[leg]  (q)    -- (n12);
  \draw[leg]  (n12)  -- node[lbl, right=3pt]
              {\cite{Kumar:2001iu,Bourget:2015cza}} (cmt);
  \draw[spec] (q)    -- (cmt);
  \draw[ours] (bae2) -- node[lbl, below]{this work:\\ $\mathcal{A}$} (cmu);
  \draw[cut]  (bae2) -- node[lbl, left]{$\times$} (cmt);

  \node[font=\small] at (2.2,-5.7) {(b) non-simply-laced $\alg$};
\end{scope}
\end{tikzpicture}
\caption{\justifying For simply-laced $\alg$ (a) the twisted and untwisted
Calogero--Moser systems coincide and the triangle closes. For non-simply-laced
$\alg$ (b) they do not: $\mathcal{A}$ lands on the untwisted system (solid
arrow), whereas $\mathcal{N}=1^\ast$ is governed by the twisted one, so the
$\mathcal{N}=4$ index cannot sit at the third vertex (dashed red). What does
remains open (dotted). Double-headed arrows denote correspondences established
in the literature; the map $\mathcal{A}$ in \eqref{eq:Ofer} is one-way, and we conjecture it to be a bijection on the preimage of the Calogero--Moser extrema.}
\label{fig:triangles}
\end{figure*}

Alongside the computational problem there is a structural one. It has been
conjectured \cite{ArabiArdehali:2019orz,Benini:2021ano} that the isolated
solutions of the $\mathcal{N}=4$ BAEs are in one-to-one correspondence with the
massive vacua of the mass-deformed $\mathcal{N}=1^\ast$ theory with the same
gauge algebra, continuous families of BAE solutions matching vacua with
unbroken $\mathfrak{u}(1)$ factors. The conjecture is natural in type $A$, where
the Hong--Liu solutions \cite{Hosseini:2016cyf,Hong:2018viz} are independent of the flavor
chemical potentials and coincide with the known $\mathcal{N}=1^\ast$ vacuum
data. But the $\mathcal{N}=1^\ast$ vacua are themselves extrema of an elliptic
superpotential, obtained for $\mathfrak{su}(N)$ in \cite{Dorey:1999sj} and
argued in \cite{Kumar:2001iu} to equal, for any gauge algebra, the potential of
the \emph{twisted} elliptic CM system, the twisting being strictly required for
non-simply-laced algebras \cite{DHoker:1998zuv}. The extrema of those twisted
systems were determined for low-rank $B$, $C$ and $D$ root systems in
\cite{Bourget:2015cza}, and the corresponding vacua counted and organized under
duality in \cite{Bourget:2015lua,Bourget:2015upj,Bourget:2016yhy}; the
resulting structure, which for $\mathfrak{so}(5)$ and $G_2$ concerns precisely
the algebras we study, differs substantially from the untwisted case.\footnote{See
also \cite{Damia:2025bla} for a notion of global variant of the twisted
Calogero--Moser system itself, matching the choice of line-operator lattice
\cite{Aharony:2013hda} with a choice of gauged translations of the integrable
system; the explicit non-simply-laced case treated there is $\mathfrak{so}(5)$, with the orbit structure of the extrema presented in their Fig.~17. The latter differs substantially from our Fig.~\ref{fig:B2CMPSL2Z}.} This is
the tension we resolve here (Fig.~\ref{fig:triangles}).

\medskip

In this note we construct a direct map from BAE solutions to CM extrema, valid
for every semisimple gauge algebra. The map is simply the limit of vanishing flavor chemical potentials,
$\Delta_a\to0$, taken by following a solution $u(\Delta_a)$ continuously as the
$\Delta_a$ are switched off. Its existence rests on an elementary observation, which appeared for the first time in
 \cite{aharony-mamroud} in the $\mathfrak{su}(N)$ case: differentiating the BAEs three times with respect to the $\Delta_a$ and then sending them to zero makes every term involving the derivatives
$\partial_a u$ of the solution cancel by parity, leaving the extremum equations
of a CM potential. We generalize this to arbitrary $\alg$ in
Section~\ref{sec:map}.

The key point, and the reason the map bears on the $\mathcal{N}=1^\ast$
conjecture, is \emph{which} CM system it produces. In the BAEs every root
enters on the same footing, with the same multiplicity and the same functional
dependence; the equations know nothing of root lengths. Consequently the map
lands, for every $\alg$, on the \emph{untwisted} CM system with equal couplings
on long and short roots. For simply-laced algebras this is immaterial, since
twisted and untwisted coincide. For non-simply-laced algebras it is not: the
map cannot reach the twisted system that governs $\mathcal{N}=1^\ast$. It
follows that the conjectured bijection between $\mathcal{N}=4$ BAE solutions
and $\mathcal{N}=1^\ast$ vacua, which holds in type $A$, cannot extend to
$B$, $C$, $F_4$ or $G_2$. What survives, we propose, is the correspondence with the untwisted system in
place of the twisted one. This is the conjecture of
Section~\ref{sec:conjecture}. We stress that this is a statement about
integrable systems: whereas the twisted system computes the vacua of
$\mathcal{N}=1^\ast$, we are not aware of a gauge theory whose vacua are
enumerated by the extrema of the untwisted system in non-simply-laced type. We
note that the untwisted non-simply-laced models arise by folding simply-laced
ones, with the long and short couplings then in a fixed ratio
\cite{Bordner:1998xs}, which is exactly the equal-coupling system our map
produces; identifying the gauge-theory counterpart of that folding, if there is
one, would be of evident interest.

We establish several properties of the map. It is well defined; it intertwines
all the symmetries of the two problems, so that orbits map to orbits; and it
sends the trivial BAE solutions, i.e. those on which some root lands on the period
lattice and which do not contribute to the index, to the poles of the CM
potential. Less expectedly, certain \emph{nontrivial} solutions also map to
poles. The image of the map is therefore a proper subset of the CM extrema
together with these poles, and we call the preimage of the CM extrema ``the CM
sector'' of the BAE solutions. Our main conjecture is that the map restricted to this sector is a
bijection.

We illustrate the map, and test the conjecture, for every rank-two semisimple algebra, classical and
exceptional. For $A_2$ and $D_2$ all isolated solutions are $\Delta_a$-independent, the
map acts as the identity, and the statement is immediate; for $A_2$ we also
verify that the continuous family of \cite{Benini:2021ano} maps onto the
continuous family of CM extrema of \cite{Bourget:2015upj}. (There are no continuous families in $D_2$.) The 
$B_2\cong C_2$ case is the first with genuinely $\Delta_a$-dependent solutions, and
the first where nontrivial solutions flow to poles: of the nine half-rational
solutions, six reach CM extrema and three, forming a complete
$\mathrm{PSL}(2,\mathbb{Z})$ orbit, do not (see Fig.~\ref{fig:B2CMPSL2Z}). For $G_2$, where we work
numerically, 15 of the 29 nontrivial solutions constitute the CM sector,
in bijection with the 15 CM extrema. Along the way we compute analytically the extrema of
the untwisted $B_2\cong C_2$ and $D_2$ CM systems, which to our knowledge have not
appeared in the literature.

\medskip

The rest of this letter is organized as follows. In Section~\ref{sec:map} we introduce
the BAEs and the untwisted CM system, define the map, and derive its
properties. In Section~\ref{sec:conjecture} we state the bijectivity conjecture,
discuss the interpretation of the non-CM sector, and spell out the relation to
the $\mathcal{N}=1^\ast$ conjecture. Section~\ref{sec:evidence} illustrates how the map works in rank two, with the derivations of the CM extrema collected in Appendix~\ref{app:CM}.

%%%%%%%%%%%%%%%%%%%%%%%%%%%%%%%%%%%%%
\section{The map}
\label{sec:map}
%%%%%%%%%%%%%%%%%%%%%%%%%%%%%%%%%%%%%

In this section we introduce the two systems and the map $\mathcal{A}$ relating them.

\subsection{Bethe Ansatz equations}
\label{sub:BAE}

For a 4D $\mathcal{N}=1$ gauge theory one can define a supersymmetric index,
the partition function on $S^1\times S^3$,
\begin{equation}\label{eq:index_def}
  \mathcal{I}=\tr\,(-1)^F e^{-\beta\{\mathcal{Q},\mathcal{Q}^\dagger\}}
  \prod_k \mathfrak{f}_k^{\,\mathcal{F}_k}\ ,
\end{equation}
counting with sign $(-1)^F$ the states annihilated by a chosen supercharge
$\mathcal{Q}$ and its conjugate \cite{Kinney:2005ej,Romelsberger:2005eg}. Here
the trace is over the Hilbert space on $S^3$, $\beta$ is proportional to the
circumference of the $S^1$, and the $\mathfrak{f}_k$ are fugacities for a
maximal set of charges $\mathcal{F}_k$ commuting with $\mathcal{Q}$: the two
angular momenta on $S^3$, and the flavor and R-symmetry charges. By standard arguments, only states
with $\{\mathcal{Q},\mathcal{Q}^\dagger\}=0$ contribute, so $\mathcal{I}$ is
independent of $\beta$ and depends only on the fugacities. When the theory flows to a CFT in the infrared, the index is invariant along
the flow and coincides with the superconformal index of the fixed point, which
counts local operators in short representations of the superconformal algebra.
It depends only on the gauge \emph{algebra} and on the matter representations
of that algebra, not on the global form of the gauge group; the global form
nevertheless has to be chosen in order to write the localization formula of \cite{Romelsberger:2005eg,Dolan:2008qi}, since
projecting onto gauge singlets requires the Haar measure of a group, and with
it a cocharacter lattice and a primitive basis thereof \cite{B2paper}.

We specialize throughout to 4D $\mathcal{N}=4$ SYM with semisimple gauge algebra
$\alg$, which in $\mathcal{N}=1$ language comprises a vector multiplet and three
adjoint chiral multiplets $X_{1,2,3}$ with superpotential
$W=\mathrm{Tr}\,X_1[X_2,X_3]$. The charges commuting with $\mathcal{Q}$ are then the two angular momenta and
two Cartan generators of the $\mathrm{SU}(3)\subset\mathrm{SU}(4)_R$ that
commutes with $\mathcal{Q}$; we write $\sigma,\tau$ for the corresponding
angular chemical potentials and $\Delta_{1,2,3}$ for those conjugate to the
three chirals, obeying $\Delta_1+\Delta_2+\Delta_3-\sigma-\tau\in\mathbb{Z}$.
We work at equal angular potentials, $\sigma=\tau\eqqcolon\omega$ (equivalently
$p=q\coloneqq e^{2\pi i \omega}$), so that the constraint reads $\Delta_3=2\omega-\Delta_1-\Delta_2$ up to
an integer, and $\mathcal{I}$ is a function of $\omega$ and $\Delta_{1,2}$
alone.

Being independent of the gauge coupling, the index localizes to a contour
integral over the maximal torus of the gauge group $G$, with
$\operatorname{Lie}G=\alg$ \cite{Romelsberger:2005eg,Dolan:2008qi}. Writing the
holonomies additively as $u\in\mathfrak{h}$ (with $\mathfrak{h}$ the Cartan subalgebra of $
\alg$), it takes the form
\begin{equation}\label{eq:int_formula}
  \mathcal{I}(\omega;\Delta_a)=\kappa_\alg\int_{[0,1]^{\rank\alg}}
  \mathcal{Z}(u;\Delta_a,\omega)\prod_{i=1}^{\rank\alg}du^i\ ,
\end{equation}
with $\kappa_\alg \coloneqq (q;q)_\infty^{2\,\mathrm{rk}(\mathfrak{g})}/|\mathcal{W}_{\mathfrak{g}}|$ and
  \begin{equation}\label{eq:Zintegrand}
  \mathcal{Z}(u;\Delta_a,\omega) \coloneqq 
  \frac{\prod_{a=1}^{3}\prod_{\rho\in\mathcal{R}}
  \tilde\Gamma\big(\langle\rho,u\rangle+\Delta_a;\omega,\omega\big)}
  {\prod_{\alpha\in\Phi}
  \tilde\Gamma\big(\langle\alpha,u\rangle;\omega,\omega\big)}\ .
\end{equation}
Here $(q;q)_\infty$ is the $q$-Pochhammer symbol, $\Phi$ the set of nonzero roots of $\alg$, $\mathcal{W}_\alg$ the Weyl group of $\alg$, $\mathcal{R}$ the set of
weights of the adjoint representation, and $\tilde\Gamma$ the elliptic gamma
function in additive variables \cite{Felder:1999vf}. The
integration domain is a fundamental cell of the cocharacter lattice
$L=\bigoplus_i\mathbb{Z}e_i$, i.e.\ the compact torus
$\mathfrak{h}_\mathbb{R}/L$.

Writing \eqref{eq:int_formula} therefore requires fixing a cocharacter lattice
$L$, between the coroot and coweight lattices, and a primitive basis
$\{e_i\}$ of it. We expand $u=\sum_iu^ie_i$ and write
$\alpha_i\coloneqq\langle\alpha,e_i\rangle$ for the components of a root in the dual
basis; the equations below depend on this choice, even though the index does
not. We refer to \cite{B2paper} for a full discussion.

Assuming the relevant equations admit only isolated solutions, the index
\eqref{eq:int_formula} can be recast as a finite sum
\cite{Closset:2017bse,Benini:2018mlo}
\begin{equation}\label{eq:BAformula}
  \mathcal{I}(\omega;\Delta_a)=\kappa_\alg
  \sum_{\hat u\,\in\,\text{BAEs}}\mathcal{Z}(\hat u;\Delta_a,\omega)\,
  H^{-1}(\hat u;\Delta_a,\omega)\ ,
\end{equation}
running over the solutions $\hat u$ of the \emph{Bethe Ansatz equations} (BAEs)
$Q_i=1$, with
$H\coloneqq\det_{ij}\big[(2\pi i)^{-1}\partial Q_i/\partial u^j\big]$ the
Jacobian of the Bethe Ansatz operators $Q_i$. For $\mathcal{N}=4$ SYM the latter
take the form
\begin{equation}\label{eq:theta1_BAE}
  Q_i=\prod_{\alpha>0}\left(-\prod_{\Delta}\frac{\theta_1(-\alpha\cdot u+\Delta)}
  {\theta_1(\alpha\cdot u+\Delta)}\right)^{\alpha_i}\ ,
\end{equation}
where $\theta_1$ is the odd Jacobi theta function
($\theta_j(x)\equiv\theta_j(x;\omega)$, $j=1,\dots,4$, at modular parameter
$\omega$), the outer product runs over the positive roots $\{\alpha>0\}$ once a set of
simple roots has been fixed, and we have set $\Delta_3=-\Delta_1-\Delta_2$,\footnote{This does not affect the BAEs $Q_i=1$,
which are modular invariant; we therefore use $\Delta_3$ and
$-\Delta_1-\Delta_2$ interchangeably.}
since $\Delta_1+\Delta_2+\Delta_3-2\omega=0 \mod
\mathbb{Z}+\omega\mathbb{Z}$, so that the inner product runs over $\Delta \coloneqq \{ \Delta_1,\Delta_2,-\Delta_1-\Delta_2\}$. This is the form in which the BAEs of
$\mathfrak{su}(N)$ SYM usually appear \cite{Benini:2018ywd}.

Two features will matter below. First, the holonomies live on the torus
$\mathbb{C}/(\mathbb{Z}+\omega\mathbb{Z})$, and solutions are identified under
the Weyl group $\mathcal{W}_\alg$, the center $Z(G)$ acting by shifts along the
two cycles, and $\mathrm{PSL}(2,\mathbb{Z})$, generated by
$T:\hat u(\omega,\Delta_a)\mapsto\hat u(\omega+1,\Delta_a)$ and
$S:\hat u(\omega,\Delta_a)\mapsto
    \omega\,\hat u\!\left(-\tfrac1\omega,-\tfrac{\Delta_a}{\omega}\right)$ on a \emph{solution} $\hat u$ of the BAEs. This 
last group acts on the modular parameter of the $S^1\times S^3$ background, and is
\emph{not} the S-duality group of the gauge theory, which acts on the holomorphic
coupling $\tau_{\mathrm{YM}}$, a parameter on which the index does not depend.
Note that $S^2\eqqcolon C$ is charge conjugation,
$C:(\omega,\Delta_a,u)\mapsto(\omega,-\Delta_a,-u)$, under which the Bethe
operators are invariant \cite[Eq.~(3.7)]{Benini:2018mlo}; for all the algebras
we consider it lies in the Weyl group, so that only $\mathrm{PSL}(2,\mathbb{Z})$
acts effectively.\footnote{This holds whenever
$-\mathbbm{1}\in\mathcal{W}_\alg$, as for $A_1$, $B_N$, $C_N$, $D_{2N}$ and
every exceptional algebra except $E_6$. For $A_N$ with $N\geq2$, $D_{2N+1}$ and
$E_6$, instead, $-\mathbbm{1}$ is the longest Weyl element composed with the
nontrivial diagram automorphism, and lies in
$\mathcal{W}_\alg\rtimes\mathrm{Out}(\alg)$ but not in $\mathcal{W}_\alg$;
there the full $\mathrm{SL}(2,\mathbb{Z})$ can act nontrivially, though in type
$A$ it squares to the identity on the Hong--Liu solutions, whose orbits are
accordingly classified by $\mathrm{PSL}(2,\mathbb{Z})$
\cite[Sec. 2]{Benini:2021ano}.} Physical solutions are orbits under the full symmetry group
\cite{B2paper}. Second, and crucially for what follows, every root enters
\eqref{eq:theta1_BAE} on the same footing: the equations are blind to root
lengths, since $\alpha$ appears only through the pairing $\alpha\cdot u$ and the
exponent $\alpha_i$.

We call a solution \emph{trivial} when some root lands on the period lattice,
$\alpha\cdot u\in\mathbb{Z}+\omega\mathbb{Z}$ for some $\alpha\in\Phi$. There the
elliptic gamma in the denominator of \eqref{eq:Zintegrand} diverges, so
$\mathcal{Z}=0$ and the solution does not contribute to
\eqref{eq:BAformula}.\footnote{Physically, such a configuration leaves a
non-abelian gauge symmetry unbroken, and the vector-multiplet measure
degenerates.}

\subsection{The untwisted elliptic Calogero--Moser system and its extremum
equations}
\label{sub:CM}

The \emph{untwisted} elliptic Calogero--Moser (CM) integrable system of type
$\alg$ \cite{Calogero:1975ii,Moser:1975qp,Olshanetsky:1981dk} is defined by the
Hamiltonian\footnote{Roots of different length may in general carry different
coupling constants, one per Weyl orbit; for our purposes we set
$g_\text{short}=g_\text{long}=1$. This uniform choice is not an idle
simplification: as will become clear in Section~\ref{sub:derivation}, it is precisely
what the map produces, and it is the reason the map cannot reach the twisted
systems relevant for non-simply-laced $\mathcal{N}=1^\ast$.}
\begin{equation}\label{eq:CMHam}
  H=\frac12\,p\cdot p+V(q)\ ,\quad V(q)\coloneqq\sum_{\alpha>0}\wp(\alpha\cdot q)\ ,
\end{equation}
where $\wp$ is the Weierstrass elliptic function. Configurations extremizing $V$ solve the \emph{extremum equations},
\begin{equation}\label{eq:CMeqn}
  \sum_{\alpha>0}\alpha_i\,\wp'(\alpha\cdot q)=0\ ,
\end{equation}
whose solutions we call \emph{CM extrema}. (They are stationary points of $V$
and need not be local minima.)

It is essential to contrast the untwisted system \eqref{eq:CMHam} with the
\emph{twisted} one, which is the system arising in the gauge/integrability
correspondence for non-simply-laced algebras \cite{DHoker:1998zuv}. The two
coincide for simply-laced $\alg$, where there is a single root length and hence
nothing to twist, and differ otherwise.\footnote{Both are integrable for any root system
\cite{Olshanetsky:1976,Olshanetsky:1981dk}; for the untwisted case a Lax pair
with spectral parameter was constructed for every simple Lie algebra, classical
and exceptional, in \cite{DHoker:1998xad}. Moreover, the massive vacua of the
$\mathcal{N}=1^\ast$ theory, which the twisted extrema compute, were counted for
every gauge algebra in \cite{Bourget:2015lua,Bourget:2015upj} by nilpotent-orbit methods.}

\subsection{The map}
\label{sub:themap}

We now define the main object of this note, a map from BAE solutions to CM
extrema:
\begin{equation}\label{eq:Ofer}
  \begin{array}{rcc}
    \mathcal{A}:\ \{\text{BAE solutions}\} & \to &
      \{\text{CM extrema}\}\cup\{\text{poles of }V\}\\[2pt]
    \hat u(\Delta_a,\omega) & \mapsto & \hat u(0,\omega)\eqqcolon q(\omega)
  \end{array}
\end{equation}
where the evaluation at $\Delta_a=0$ is understood as tracking the solution
continuously as the $\Delta_a$ are turned off, so as to avoid branch cuts. The following properties hold:
\begin{itemize}
  \item[\emph{i)}] \otherlabel{map:i}{\emph{i)}} the map is well defined;
  \item[\emph{ii)}] \otherlabel{map:ii}{\emph{ii)}} trivial BAE solutions map to poles of $V$;
  \item[\emph{iii)}] \otherlabel{map:iii}{\emph{iii)}} nontrivial BAE solutions may also map to poles;
  \item[\emph{iv)}] \otherlabel{map:iv}{\emph{iv)}} all nontrivial $\Delta_a$-independent solutions map to CM extrema;
  \item[\emph{v)}] \otherlabel{map:v}{\emph{v)}} the map preserves the symmetries of the equations, both the gauge ones
  (torus, Weyl and center invariance) and the $\mathrm{PSL}(2,\mathbb{Z})$
  action on solutions and extrema.
\end{itemize}
We call the preimage of the set of CM extrema the \emph{CM sector} of BAE
solutions.

\subsection{Derivation of the map and its properties}
\label{sub:derivation}

To prove that the map is well defined (property \ref{map:i}) we generalize an unpublished argument \cite{aharony-mamroud} of
O.~Aharony, communicated to us by O.~Mamroud, relating the $\mathfrak{su}(N)$
BAEs to the extremum equations of the elliptic CM system of type $A$, a case in
which twisted and untwisted potentials coincide, as they do precisely when
$\alg$ is simply-laced.

We use repeatedly the identity
\begin{equation}\label{eq:wpident}
  \wp(x;\omega)=-\partial_x^2\log\theta_1(x;\omega)+c(\omega)\ ,
\end{equation}
where $c(\omega)$ is independent of $x$,\footnote{Explicitly,
$c(\omega)=\tfrac13\,\theta_1'''(0;\omega)/\theta_1'(0;\omega)$. Both sides of
\eqref{eq:wpident} are elliptic in $x$ with a double pole at the origin, so they
agree up to an additive constant; $c$ is fixed by the normalization that the
Laurent expansion of $\wp$ at $x=0$ has vanishing constant term. Its precise
form will play no role below.} and write $\Theta_1(x)\coloneqq\partial_x\log\theta_1(x)$, so that $\Theta_1'=-\wp+c$
and $\Theta_1''=-\wp'$.\footnote{Throughout, a prime denotes differentiation with
respect to the argument of the elliptic function, not with respect to the
chemical potentials, for which we reserve $\partial_a\coloneqq\partial/\partial\Delta_a$.} The argument is comprised of two steps.

\subsubsection{Differentiating three times}

Define the single-root block
\begin{equation}\label{eq:Gblock}
  G_\alpha(\Delta)\coloneqq\sum_{a=1}^{3}
  \log\frac{\theta_1(\alpha\cdot \hat u+\Delta_a)}{\theta_1(\alpha\cdot \hat u-\Delta_a)}\ ,
  \quad F_i\coloneqq\sum_{\alpha>0}\alpha_i\,G_\alpha\ ,
\end{equation}
so that $F_i$ agrees with $\log Q_i$ up to an overall sign and an additive
constant, neither of which affects the vanishing of the third derivative below.
Impose the constraint on $\Delta_a$ of Section~\ref{sub:BAE}. It is convenient
to further shift $\Delta_3\to\Delta_3+2\omega$, so that the three potentials sum to zero
and all three summands of \eqref{eq:Gblock} are treated on the same footing.\footnote{We pay no cost for shifting the $\Delta_a$ as done, since $F_i \propto \log Q_i $ is invariant under torus period shifts when evaluated on a BAE solution.} Call $u\equiv u(\Delta)$; the first derivative of $G_\alpha$ reads
%\begin{widetext}
\begin{eqnarray}\label{eq:firstderiv}
  \partial_1G_\alpha= \sum_{a=1}^3&\left[\Theta_1(\alpha\cdot \hat u+\Delta_a)\,
    \left(\partial_1\Delta_a+\alpha\cdot\partial_1 \hat u\right)+\right.\nonumber \\
   &\left.-\Theta_1(\alpha\cdot \hat u-\Delta_a)\,
    \left(\partial_1\Delta_a-\alpha\cdot\partial_1 \hat u\right)\right]\ ,
\end{eqnarray}
%\end{widetext}
with $\partial_1\Delta_1=+1$, $\partial_1\Delta_2=0$, $\partial_1\Delta_3=-1$.
The second derivative $\partial_2\partial_1G_\alpha$ brings down
$\Theta_1'=-\wp+c$ and $\partial_2\partial_1u$; the third derivative
$\partial_1\partial_2\partial_1G_\alpha$ brings down $\Theta_1''=-\wp'$ and
$\wp'$. For brevity we do not write out these straightforward intermediate
expressions.

\subsubsection{Setting \texorpdfstring{$\Delta_a\to0$}{Delta to 0} and collecting parities}

At $\Delta_1=\Delta_2=0$ (hence $\Delta_3=0$ after the shift on the torus) all arguments collapse to
$\pm\,\alpha\cdot \hat u$, and we use $\wp(\pm w)=\wp(w)$, $\wp'(\pm w)=\pm\wp'(w)$
and $\Theta_1(\pm w)=\pm\Theta_1(w)$. Every term carrying a factor of
$\alpha\cdot\partial_a \hat u$ or $\alpha\cdot\partial_a\partial_b \hat u$ multiplies either
$\wp(w)$ (even) or $\Theta_1(w)$ (odd) in a $\pm$-symmetric combination that
cancels. The lone survivor is the $\wp'$ piece, one per root:
\begin{equation}\label{eq:A2der}
  \partial_1\partial_2\partial_1 G_\alpha\big|_{\Delta_a=0}
  =2\,\wp'(\alpha\cdot \hat u (\Delta_a \to 0))\ .
\end{equation}
Note that $c(\omega)$, being independent of $x$, is annihilated by the third
derivative and hence does not appear.

Applying the repeated derivative to the BAEs \eqref{eq:theta1_BAE} in the form
$\log Q_i=2\pi i k$, where the integer $k$ records the branch of the logarithm
and is annihilated by every $\partial_a$, we therefore obtain
\begin{align}\label{eq:A3}
  0&=\tfrac12\,\partial_1\partial_2\partial_1F_i\big|_{\Delta_a=0}
   =\tfrac12\sum_{\alpha>0}\alpha_i\,
    \partial_1\partial_2\partial_1G_\alpha\big|_{\Delta_a=0}\notag\\
   &=\sum_{\alpha>0}\alpha_i\,\wp'\left(\alpha\cdot \hat u(\Delta_a\to 0)\right)\ ,
\end{align}
which are precisely the untwisted CM extremum equations \eqref{eq:CMeqn} for
the root system of $\alg$ under the identification $q(\omega)= \hat u(0)$.

Note that this argument \emph{assumes} $\wp'(\alpha\cdot\hat u(\Delta_a\to0))$ to be finite for every $\alpha$. Failure of that assumption is precisely the loophole through which some nontrivial BAE solutions flow to poles of $V$; see property~\ref{map:iii}.

\subsubsection{The remaining properties \ref{map:ii}--\ref{map:v}}
\label{subsub:otherprops}

To show that a trivial BAE solution must map to a pole of $V$ (property
\ref{map:ii}), note first that the zero locus of $\mathcal{Z}$ and the polar
locus of $V$ coincide. For generic $\Delta_a$,
\begin{align}
  \mathcal{Z}(\hat u;\Delta_a,\omega)=0 &\iff \exists\,\alpha\in\Phi:\
    \alpha\cdot \hat u\in\mathbb{Z}+\omega\mathbb{Z}\ ,\\
  V(q;\omega)=\infty &\iff \exists\,\alpha\in\Phi:\
    \alpha\cdot q\in\mathbb{Z}+\omega\mathbb{Z}\ ,
\end{align}
On the $\mathcal{Z}$ side we assume in addition that $\hat u$ is a BAE solution, which
costs no generality in the context of the map.\footnote{This assumption is necessary in order to have full double periodicity of $\mathcal{Z}$ on the torus.} The two conditions are then
equivalent: since $\hat u$ depends continuously on the $\Delta_a$
along the branch being tracked (a statement about the \emph{deformation} of a
given solution, not to be confused with membership in a continuous family at
fixed $\Delta_a$), the limit must satisfy
$\alpha\cdot \hat u(\Delta_a\to0) = \alpha \cdot q \in\mathbb{Z}+\omega\mathbb{Z}$ for the same
$\alpha$.

For a case where the third property \ref{map:iii} holds, we refer to $B_2\cong C_2$
and $G_2$ in Sections~\ref{sub:B2} and~\ref{sub:G2} respectively. The
fourth \ref{map:iv} is an immediate corollary of the first two. Preservation of the symmetries (property \ref{map:v}) is immediate for the gauge ones: torus shifts,
Weyl reflections and center shifts all act on $\hat u$ alone, and the map is the
identity on $\hat u$, so equivariance is automatic. The
$\mathrm{PSL}(2,\mathbb{Z})$ case has more content. Under
$S$, $\wp$ transforms with weight two and $\wp'$ with weight three,
$\wp'(q/\omega;-1/\omega)=\omega^3\,\wp'(q;\omega)$, so \eqref{eq:CMeqn} is
covariant provided the weight is the same for every root. That is the case
here, since all roots enter \eqref{eq:CMHam} with a single common coupling; it
would fail for a twisted potential, whose couplings are distributed over the
half-periods of the torus.

\subsection{The map on continuous families}
\label{sub:continuous}

The map applies in the same way to continuous families of solutions.
This is an important point, since it is by now known that such families exist
generically for algebras of rank $>2$, on both the BAE and the CM extremum
side. In this note we do not dwell on the continuous counterpart of the map,
because at rank two a continuous family exists only for $A_2$; we treat that
case in Appendix~\ref{subsub:A2ContApp}.

Concretely, a family is a $k$-dimensional set of solutions at fixed
$\Delta_a$; each of its points carries its own $\Delta_a$-dependence, and the
map acts pointwise, tracking each as the potentials are sent to zero.

%%%%%%%%%%%%%%%%%%%%%%%%%%%%%%%%%%%%%
\section{The bijectivity conjecture}
\label{sec:conjecture}
%%%%%%%%%%%%%%%%%%%%%%%%%%%%%%%%%%%%%

In the previous section we established some basic properties of the map.
We now put forward a conjectural statement relating the physics of the
superconformal index of $\mathcal{N}=4$ SYM to that of integrable models. It
can be stated succinctly:
\begin{gather*}
  \textbf{The map is a bijection on the}\\
  \textbf{CM sector of BAE solutions.}
\end{gather*}
In particular, each CM extremum deforms to a unique, in general
$\Delta_a$-dependent, BAE solution. The conjecture rests on the empirical
evidence gathered in Section~\ref{sec:evidence} from all $\mathcal{N}=4$ SYM theories
with rank-two semisimple gauge algebra. We state it for isolated solutions; the
$A_2$ family of Appendix~\ref{subsub:A2ContApp} suggests it extends to
continuous ones, but a single example is thin evidence and we do not claim it.

\subsection{Interpretation of the non-CM sector}
\label{sub:sectors}

An immediate question is what interpretation to give to those nontrivial BAE
solutions that map to poles of $V$. We do not have a complete answer, but two
remarks are in order.

First, it is worth distinguishing two things that degenerate as
$\Delta_a\to0$. The limit is the unrefinement of the index,
$y_a=e^{2\pi i\Delta_a}\to1$, and it is a familiar one. It was already noticed
in \cite[Sec.~2]{Spiridonov:2010qv}, whose ``balancing condition'' reads
$y_1y_2y_3=pq$ in our conventions, equivalently
$\Delta_1+\Delta_2+\Delta_3-\sigma-\tau\in \mathbb{Z}$ in additive variables. There
it is observed that sending a single $y_a\to1$ cancels the elliptic gamma
functions in the denominator of \eqref{eq:Zintegrand} against those in the
numerator; the remaining pair then satisfies the reflection property
$\tilde\Gamma(\Delta_1,\Delta_2;\sigma,\tau)=1$, valid whenever
$y_1y_2=pq$, and the integrand becomes \emph{identically} equal to one. All the
singular behavior thus resides in the $u$-independent prefactor, in our
notation $\kappa_\alg$ times the contribution of the $\rank\alg$ zero
weights of the adjoint, and the limit says nothing about individual holonomy
configurations.\footnote{Indeed, \cite{Spiridonov:2010qv} uses it precisely as a check
that the ratio of the indices of the S-dual 4D $B_N$ and
$C_N$ SYM theories tends to one.}

Note also that the fully symmetric limit $\Delta_1=\Delta_2=\Delta_3=0$ is
\emph{incompatible} with the balancing condition at generic $\omega$ (with, as before, $\omega\coloneqq\sigma=\tau$). What we
take instead is $\Delta_{1,2}\to0$ with $\Delta_3\to2\omega$. On the torus
$\mathbb{C}/(\mathbb{Z}+\omega\mathbb{Z})$ these are the same point, but the
index is not a function on that torus: the elliptic gamma functions are
quasi-periodic rather than periodic, so shifting an argument by $\omega$
produces an explicit factor. The two limits are therefore genuinely distinct
for the integrand, though, as shown in Section~\ref{sub:derivation}, the
difference does not affect the map.

The degeneration relevant to us is a different, $u$-dependent one: $V$ diverges
exactly where some root lands on the period lattice of the torus. The map is defined on
\emph{solutions} $\hat u(\Delta_a)$, which continue to exist and to have finite
limits as $\Delta_a\to0$ irrespective of the fate of the integrand; whether such
a limit is an CM extremum or a pole of $V$ is a property of the solution, not of
the index of $\mathcal{N}=4$.

Second, the limit itself does have a physical interpretation, even if the
solutions flowing to poles do not yet. It is an instance of the vacuum
expectation value (VEV) deformations of \cite[Sec.~7.3]{Rastelli:2016tbz}, in
which switching off the fugacities for the symmetries broken by a VEV produces a
pole of the index. The divergence is quantitatively of the right kind: since
$\tilde\Gamma$ has a simple pole at vanishing argument and a simple zero at
$2\omega$, the zero-weight prefactor contributes
\begin{equation}\label{eq:polecount}
    \tilde\Gamma(0)^{2\rank\alg}\,\tilde\Gamma(2\omega)^{\rank\alg}
    \ \sim\ (\text{pole})^{\rank\alg}\ ,
\end{equation}
a pole of order exactly $\rank\alg$, matching the $\rank\alg$ moduli acquiring a
VEV. The operators in question are the independent Casimirs built from the
Cartan-valued components of the $X_a$, which parameterize the Coulomb branch of
the $\mathcal{N}=4$ theory; unlike the Higgs-branch case, in which the residue
computes the index of a new infrared fixed point, here one moves along that
branch rather than flowing to a different CFT.\footnote{We thank
S.~S.~Razamat for explaining this to us, and for pointing out that in
$\mathcal{N}=2$ language our limit is the $\alpha=2$ member of the family of ``generalized Schur" limits
$pq/t=p^\alpha$ of \cite{Zeev:2026alb}, of which the original Schur limit is
$\alpha=1$; the deformations of Coulomb-branch operators studied there
correspond to $\alpha=0$. Related limits for $\mathcal{N}=2$ theories are
discussed in \cite{Deb:2025ypl,Cecotti:2015lab}.} What we lack is therefore not
an interpretation of the limit but a criterion distinguishing, among the BAE
solutions, those that reach CM extrema from those that do not. We observe that
in the $B_2\cong C_2$ case the three of the latter form a complete
$\mathrm{PSL}(2,\mathbb{Z})$ orbit, so that the CM sector is itself a union of
orbits, consistent with the equivariance of the map
(see Fig.~\ref{fig:B2CMPSL2Z}).

\subsection{Relation to the \texorpdfstring{$\mathcal{N}=1^\ast$}{N=1*}
conjecture}
\label{sub:N1star}

Our conjecture is related to an earlier one linking BAE solutions to vacua of
$\mathcal{N}=1^\ast$, first put forward in \cite{ArabiArdehali:2019orz} and
sharpened in \cite{Benini:2021ano}, where it is stated as a one-to-one
correspondence between branches of solutions of the ``reduced BAEs'' (which, as shown in \cite{B2paper}, amount to a particular choice of global form and primitive basis in \eqref{eq:int_formula}--\eqref{eq:Zintegrand} for type $A$) and vacua of
the $\mathcal{N}=1^\ast$ theory on $\mathbb{R}^{3,1}$: isolated solutions
correspond to massive vacua, and complex $k$-dimensional families of solutions
to Coulomb vacua with $k$ massless photons.\footnote{The conjecture as originally formulated in \cite[Sec. 2.1.1]{ArabiArdehali:2019orz}
concerns the theory compactified on $\mathbb{R}^3\times S^1$, whereas
\cite{Benini:2021ano} argues that the uncompactified one is the right comparison.
The argument is a count of vacua: on $\mathbb{R}^{2,1}\times S^1$ certain gapped
vacua split, owing to a residual discrete gauge symmetry, so that e.g.
$\mathfrak{su}(2)$ $\mathcal{N}=1^\ast$ has three vacua on $\mathbb{R}^{3,1}$
but four on $\mathbb{R}^{2,1}\times S^1$, against three isolated solutions of
the reduced $\mathcal{N}=4$ BAEs contributing to the index \cite[footnote~5]{Benini:2021ano}.}

For type $A$ the two conjectures are equivalent, provided all \emph{isolated} solutions are of Hong--Liu type \cite{Hong:2018viz}. These are independent
of $\Delta_a$, and hence all lie in the CM sector. The equivalence then follows
from that of $\mathcal{N}=1^\ast$ with the twisted CM system
\cite{Dorey:1999sj,DHoker:1998zuv,Kumar:2001iu}, which for simply-laced algebras coincides
with the untwisted one.

\begin{figure}[htb!]
  \centering
  \includegraphics[scale=0.53]{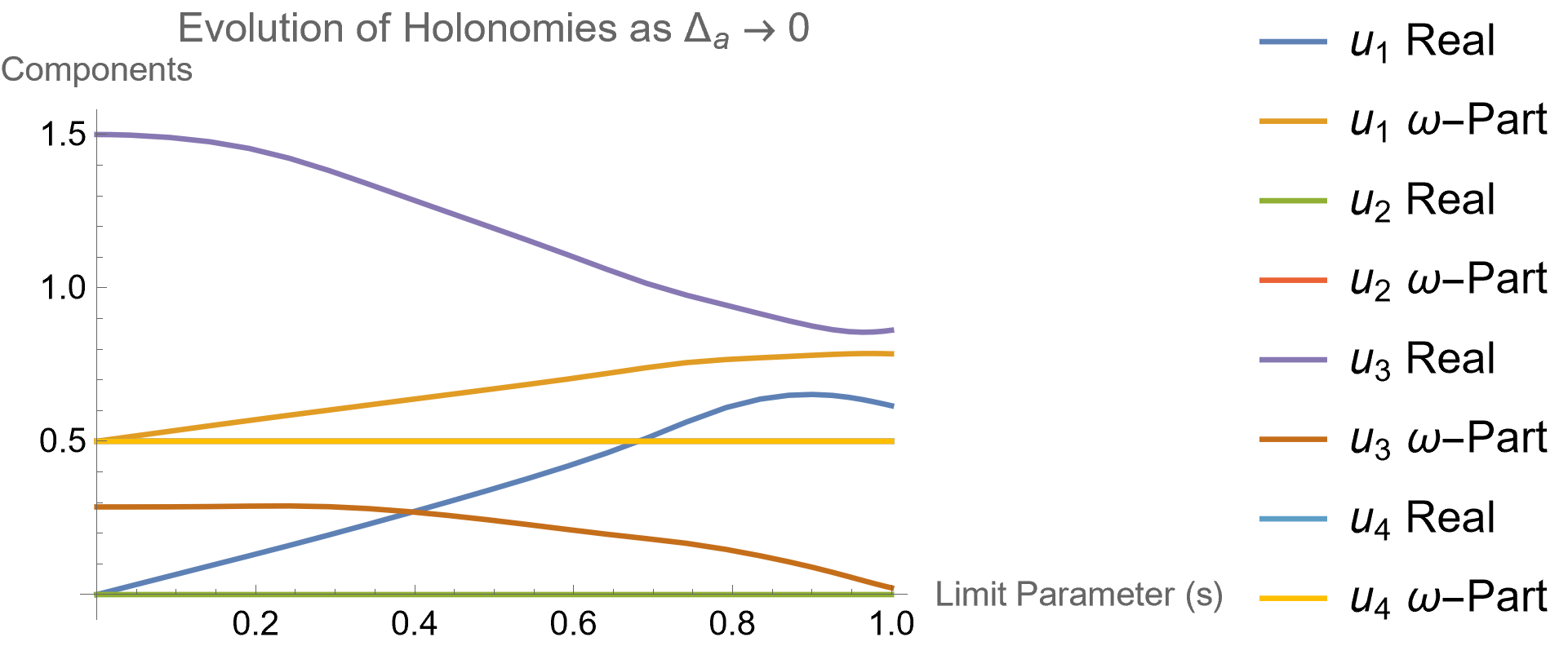}
  \caption{\justifying Tracking one $D_4$ solution, found numerically, as $\Delta_a\to0$ (the origin of the horizontal axis). The
starting values are generic, $\Delta_1=0.421+0.411i$ and $\Delta_2=0.34+0.33i$,
while $\omega$ is held fixed at the generic value $\omega=1.82i$. The solution $\hat{u}$
is ``half-rational'' in the language of Section~\ref{sub:B2}: in the chosen Weyl
representative $\hat u=\big(u^1_\ast(\Delta_a),\tfrac{\omega}{2},
u^3_\ast(\Delta_a),\tfrac{1+\omega}{2}\big)\to
\big(\tfrac{\omega}{2},\tfrac{\omega}{2},\tfrac{\omega}{2}+0.285762,
\tfrac{1+\omega}{2}\big)$ as $\Delta_a\to 0$, with $u^{1,3}_\ast(\Delta_a)$ having irrational real
and $\omega$-components. In the limit the first component reaches a half-period,
so that $\alpha\cdot\hat u(\Delta_a\to0)=0 \mod \mathbb{Z}+\omega \mathbb{Z}$ for the roots $\alpha=(1,\pm 1,0,0)$ in the orthogonal basis, and the solution therefore maps to a pole of $V$. The plots and numerical details of the analysis of this and a few other numerical $D_4$ solutions can be found in the ancillary \textsc{Mathematica} notebook included with the arXiv submission.}
  \label{fig:D4evidence}
\end{figure}

For $D_N$ and $E_{678}$ the two conjectures are likewise equivalent, $D$ and $E$ being simply-laced.
Beyond rank three, however, we expect nontrivial $\Delta_a$-dependent solutions
mapping to poles of $V$, in which case the CM sector would be a proper subset of
the solution set even in the simply-laced case.\footnote{In type $D_3 \cong A_3$, the $\mathcal{N}=1^\ast$ conjecture also implies that the $\Delta_a$-independent Hong--Liu solutions exhaust the list of isolated solutions.} Indeed, numerical evidence shows that, already in $D_4$, there exist $\Delta_a$-dependent solutions which map to poles of $V$ as $\Delta_a \to 0$; see Fig.~\ref{fig:D4evidence}.

For non-simply-laced algebras such as $B_2\cong C_2$ and $G_2$, discussed below,
the two conjectures diverge completely: there $\mathcal{N}=1^\ast$ corresponds
only to the twisted CM system, whose extremum structure 
is entirely different from that of the untwisted system we use. For such algebras the BAE solutions therefore cannot be in
bijection with the vacua of $\mathcal{N}=1^\ast$; what they are in bijection
with, we conjecture, is the extrema of the untwisted system. We are not aware
of a gauge theory whose vacua these enumerate, nor, to the best of our
knowledge, have they been studied in the integrable-systems literature, where
attention has focused on the twisted models.

A further contrast is that the duality group acting on our extrema is the full
$\mathrm{PSL}(2,\mathbb{Z})$, whereas those of the twisted systems organize into
representations of Hecke groups \cite{Bourget:2015cza}. This is as expected: the
twisted potentials distribute their couplings over the half-periods of the
torus, which modular transformations permute, whereas \eqref{eq:CMHam} has a
single coupling and the uniform covariance
$\wp'(q/\omega;-1/\omega)=\omega^3\wp'(q;\omega)$ across all roots. Both
statements concern the modular parameter of the integrable system, and should
not be conflated with the S-duality group of the gauge theory, which acts on the
holomorphic coupling $\tau_{\mathrm{YM}}$ and is likewise a Hecke group for
non-simply-laced algebras \cite{Argyres:2006qr}, but on which the index does
\emph{not} depend.

%%%%%%%%%%%%%%%%%%%%%%%%%%%%%%%%%%%%%
\section{Rank two}
\label{sec:evidence}
%%%%%%%%%%%%%%%%%%%%%%%%%%%%%%%%%%%%%

We now compare the sets of BAE solutions with the extrema of the untwisted CM
system for the rank-two algebras.\footnote{Rank one is trivial: there is a unique simple algebra, $A_1\cong B_1\cong C_1$,
with a single positive root, so twisted and untwisted potentials coincide and
the CM extrema are the three half-periods, the zeros of $\wp'$. These are
precisely the $A_1$ BAE solutions contributing to the index \cite{ArabiArdehali:2019orz,Benini:2021ano,Lezcano:2021qbj}, $\hat u=\tfrac12,\tfrac\omega2,\tfrac{1+\omega}{2}$, which are $\Delta_a$-independent, so the map acts as the identity and is trivially a bijection.} This serves two purposes: it illustrates the
map, established in Section~\ref{sec:map} for arbitrary $\alg$, showing in particular
how the different classes of BAE solutions are distributed between CM extrema
and poles of $V$; and it provides the evidence for the bijectivity conjecture of
Section~\ref{sec:conjecture}, which is not proven. For detailed derivations of the BAE
solutions we refer to \cite{B2paper}, and to
\cite{Benini:2021ano,Lezcano:2021qbj} for $A_2$. On the CM side we have found no
tabulation of extrema of the untwisted system in the literature, except for
$A_2$, where twisted and untwisted coincide and the extrema are given in
\cite{Bourget:2015upj}; we therefore present brief analytic derivations for
$D_2$ and $B_2\cong C_2$ in Appendix~\ref{app:CM}, and numerical results for $G_2$. Throughout we use, for each algebra, the same basis of the Cartan subalgebra as
in \cite{B2paper}: for the classical cases the orthogonal one, corresponding to the global forms
$\mathrm{PSU}(3)$ for $A_2$, $\mathrm{SO}(4)$ for $D_2$, and $\mathrm{SO}(5)$
for $B_2\cong C_2$, and for $G_2$, which is centerless and admits a single global
form, the unipotent basis. Neither the index nor the map depends on this
choice: a change of basis acts on both sides simultaneously.

\subsection{\texorpdfstring{$A_2=\mathfrak{su}(3)$}{A2=su(3)}}
\label{sub:A2}

This is the case in which both sides are already known (see Fig.~\ref{fig:triangles}), and we use it to fix
conventions and to illustrate the map $\mathcal{A}$ on a continuous family.

In terms of $\mathfrak{u}(3)$ holonomies $u_1,u_2,u_3$ subject to
$u_1+u_2+u_3=0$ mod $\mathbb{Z}\oplus\omega\mathbb{Z}$, the $\mathfrak{su}(3)$
BAEs read, in the ``reduced'' form of \cite{Benini:2021ano},
\begin{align}
  Q_1&=\prod_{\Delta}\frac{\theta_1(u_2-u_1+\Delta)\,\theta_1(u_3-u_1+\Delta)}
  {\theta_1(u_1-u_2+\Delta)\,\theta_1(u_1-u_3+\Delta)}=1\ ,\\
  Q_2&=\prod_{\Delta}\frac{\theta_1(u_1-u_2+\Delta)\,\theta_1(u_3-u_2+\Delta)}
  {\theta_1(u_2-u_1+\Delta)\,\theta_1(u_2-u_3+\Delta)}=1\ ,
\end{align}
the product running over
$\Delta\in\{\Delta_1,\Delta_2,-\Delta_1-\Delta_2\}$. The CM potential in
variables obeying the same constraint is
\begin{equation}\label{eq:A2pot}
  V(q_1,q_2,q_3)=\wp(q_1-q_2)+\wp(q_2-q_3)+\wp(q_3-q_1)\ .
\end{equation}

The isolated BAE solutions $(u_1,u_2,u_3)$, the Hong--Liu solutions \cite{Hong:2018viz}, are:
\begin{equation}
\label{eq:A2HL}
  \left\{\left(0,\tfrac13,\tfrac23\right),
  \left(0,\tfrac{\omega}{3},\tfrac{2\omega}{3}\right),
  \left(0,\tfrac{1+\omega}{3},\tfrac{2+2\omega}{3}\right),
  \left(0,\tfrac{1+2\omega}{3},\tfrac{2+\omega}{3}\right)\right\}\ .
\end{equation}
Being $\Delta_a$-independent, they are fixed by the map, which on them
reduces to the identity. They are precisely the (un)twisted $A_2$ CM extrema
quoted in \cite{Bourget:2015upj}. Hence all isolated $A_2$ BAE solutions lie in
the CM sector, and the conjecture is explicitly satisfied.

Benini and Rizi also showed \cite{Benini:2021ano} that the $A_2$ BAEs admit a
one-dimensional continuous family of solutions, which we rewrite in Jacobi form as the equation
\begin{equation}\label{eq:A2BAEcont}
  \cn(2Ku_{21})\,\cn(2Ku_{31})\,\cn(2Ku_{23})=D\ .
\end{equation}
Here $u_{ij}\coloneqq u_1-u_j$; $\sn$, $\cn$, $\dn$ are the Jacobi elliptic functions of elliptic parameter
$m\equiv m(\omega)$, related to the theta functions by
$m=\theta_2^4(0)/\theta_3^4(0)$; $K=K(m)$ is the complete elliptic integral
of the first kind, which sets their quarter-period; and 
\begin{subequations}\label{eq:D}
\begin{eqnarray}
  D&\coloneqq&\Big(\frac{1-m}{m}\Big)^{3/4}\frac{1}{C}\ , \\
  C&\coloneqq&\frac{\theta_4(\Delta_1)\,\theta_4(\Delta_2)\,\theta_4(\Delta_1+\Delta_2)}
  {\theta_2(\Delta_1)\,\theta_2(\Delta_2)\,\theta_2(\Delta_1+\Delta_2)}\ .  
\end{eqnarray}
\end{subequations}
Applying the map here is less immediate, but leads precisely to the continuous
family \eqref{eq:A2CMcont} of $A_2$ CM extrema derived in
\cite[Sec. 4.2]{Bourget:2015upj}. We give the details in
Appendix~\ref{subsub:A2ContApp}.

\subsection{\texorpdfstring{$D_2=\mathfrak{so}(4) \cong \mathfrak{su}(2)\oplus \mathfrak{su}(2)=A_1\oplus A_1$}{D2=so(4)=su(2)+su(2)=A1+A1}}
\label{sub:D2}

The case of $D_2$ is even simpler than $A_2$: there are no continuous families,
and all isolated solutions are $\Delta_a$-independent. All BAE solutions
therefore lie in the CM sector, the map once again reduces to the identity, and it
is trivially bijective.

For completeness, the $D_2$ BAEs in the orthogonal basis are $Q_1=Q_2=1$, with
\begin{subequations}
\begin{eqnarray}
  Q_1&=&\prod_{\Delta}
  \frac{\theta_1(-u_1-u_2+\Delta)\,\theta_1(-u_1+u_2+\Delta)}
  {\theta_1(u_1+u_2+\Delta)\,\theta_1(u_1-u_2+\Delta)}\ ,\\
  Q_2&=&\prod_{\Delta}
  \frac{\theta_1(-u_1-u_2+\Delta)\,\theta_1(u_1-u_2+\Delta)}
  {\theta_1(u_1+u_2+\Delta)\,\theta_1(-u_1+u_2+\Delta)}\ ,
\end{eqnarray}
\end{subequations}
and the CM potential is
\begin{equation}\label{eq:D2pot}
  V(q_1,q_2)=\wp(q_1-q_2)+\wp(q_1+q_2)\ .
\end{equation}
Solutions and extrema alike are listed in \eqref{eq:D2sols}. The derivation of
the CM extrema in Appendix~\ref{sub:D2CM} closely parallels that of the BAE
solutions given in \cite{B2paper}.

\subsection{\texorpdfstring{$B_2 =\mathfrak{so}(5)\cong \mathfrak{usp}(4)= C_2$}{B2=so(5)=usp(4)=C2}}
\label{sub:B2}

The $B_2$ BAEs take the explicit form
\begin{widetext}
\begin{subequations}\label{eq:BAE_B2}
\begin{eqnarray}
  Q_1&=&\prod_{\Delta}\frac{\theta_1(-u_1-u_2+\Delta)\,\theta_1(-u_1+\Delta)\,
  \theta_1(-u_1+u_2+\Delta)}
  {\theta_1(u_1+u_2+\Delta)\,\theta_1(u_1+\Delta)\,\theta_1(u_1-u_2+\Delta)}=1\ ,\\
  Q_2&=&\prod_{\Delta}\frac{\theta_1(-u_1-u_2+\Delta)\,\theta_1(-u_2+\Delta)\,
  \theta_1(u_1-u_2+\Delta)}
  {\theta_1(u_1+u_2+\Delta)\,\theta_1(u_2+\Delta)\,\theta_1(-u_1+u_2+\Delta)}=1\ ,
\end{eqnarray}
\end{subequations}
\end{widetext}
and the untwisted CM potential is
\begin{equation}\label{eq:B2pot}
  V(q_1,q_2)=\wp(q_1-q_2)+\wp(q_1+q_2)+\wp(q_1)+\wp(q_2)\ .
\end{equation}
The BAEs were solved for the first time in \cite{B2paper}, and their solutions fall into three
categories; the untwisted $B_2$ CM extrema are derived in
Appendix~\ref{sub:CMB2}. We track the map on each category in turn.

\paragraph{Fully rational solutions.}
The three simplest BAE solutions are fully rational, with Weyl representatives
\begin{equation}\label{eq:B2fullyrat}
  (u_1,u_2) = \left\{\left(\tfrac{1+\omega}{2},\tfrac{\omega}{2}\right),
  \left(\tfrac{\omega}{2},\tfrac12\right),
  \left(\tfrac{1+\omega}{2},\tfrac12\right)\right\}\ .
\end{equation}
Being $\Delta_a$-independent, they are fixed by the map, and coincide with the
\emph{fully rational extrema} \eqref{eq:B2CMfullyrational}.

\paragraph{Half-pinned solutions.}
The half-pinned solutions, so called because they satisfy
$u_2-u_1\in\tfrac12\mathbb{Z}+\tfrac{\omega}{2}\mathbb{Z}$, are the pairs
$(u_1,u_2)$ with $u_2=u_1+x$ and, in terms of $D$ from \eqref{eq:D},
\begin{subequations}\label{eq:B2halfpinnedBAE}
\begin{eqnarray}
  u_1&=&\tfrac{1}{4K}\sn^{-1}\sqrt{\tfrac{D(mD+2(1-m))}{(mD+1-m)^2}}\ , \quad  x=\tfrac12\ ,\\
  u_1&=&\tfrac{1}{4K}\sn^{-1}\sqrt{\tfrac{2mD-m+1}{m\left(1-m(1-D)^2\right)}}\ , \quad  x=\tfrac{\omega}{2}\ ,\\
  u_1&=&\tfrac{1}{4K}\sn^{-1}\sqrt{1-\left(\tfrac{m-1}{mD}\right)^{2}}\ , \quad x=\tfrac{1+\omega}{2}\ .
\end{eqnarray}
\end{subequations}
Since $\theta_4(0)/\theta_2(0)=\big((1-m)/m\big)^{1/4}$, the map sends $D\to1$.
Evaluating \eqref{eq:B2halfpinnedBAE} there gives
\begin{subequations}\label{eq:B2halfpinnedimage}
\begin{eqnarray}
  q_1&=&\tfrac{1}{4K}\sn^{-1}\left(\sqrt{2-m},m\right)\ , \quad x=\tfrac12\ ,\\
  q_1&=&\tfrac{1}{4K}\sn^{-1}\left(\sqrt{\tfrac{1+m}{m}},m\right)\ , \quad x=\tfrac{\omega}{2}\ ,\\
  q_1&=&\tfrac{1}{4K}\sn^{-1}\left(\tfrac{\sqrt{2m-1}}{m},m\right)\ ,\quad x=\tfrac{1+\omega}{2}\ ,
\end{eqnarray}
\end{subequations}
with $q_2=q_1+x$, which are exactly the \emph{half-pinned CM extrema}
\eqref{eq:B2CMhalfpinned}.

\begin{figure*}[htb!]
  \centering
  \includegraphics[width=\textwidth]{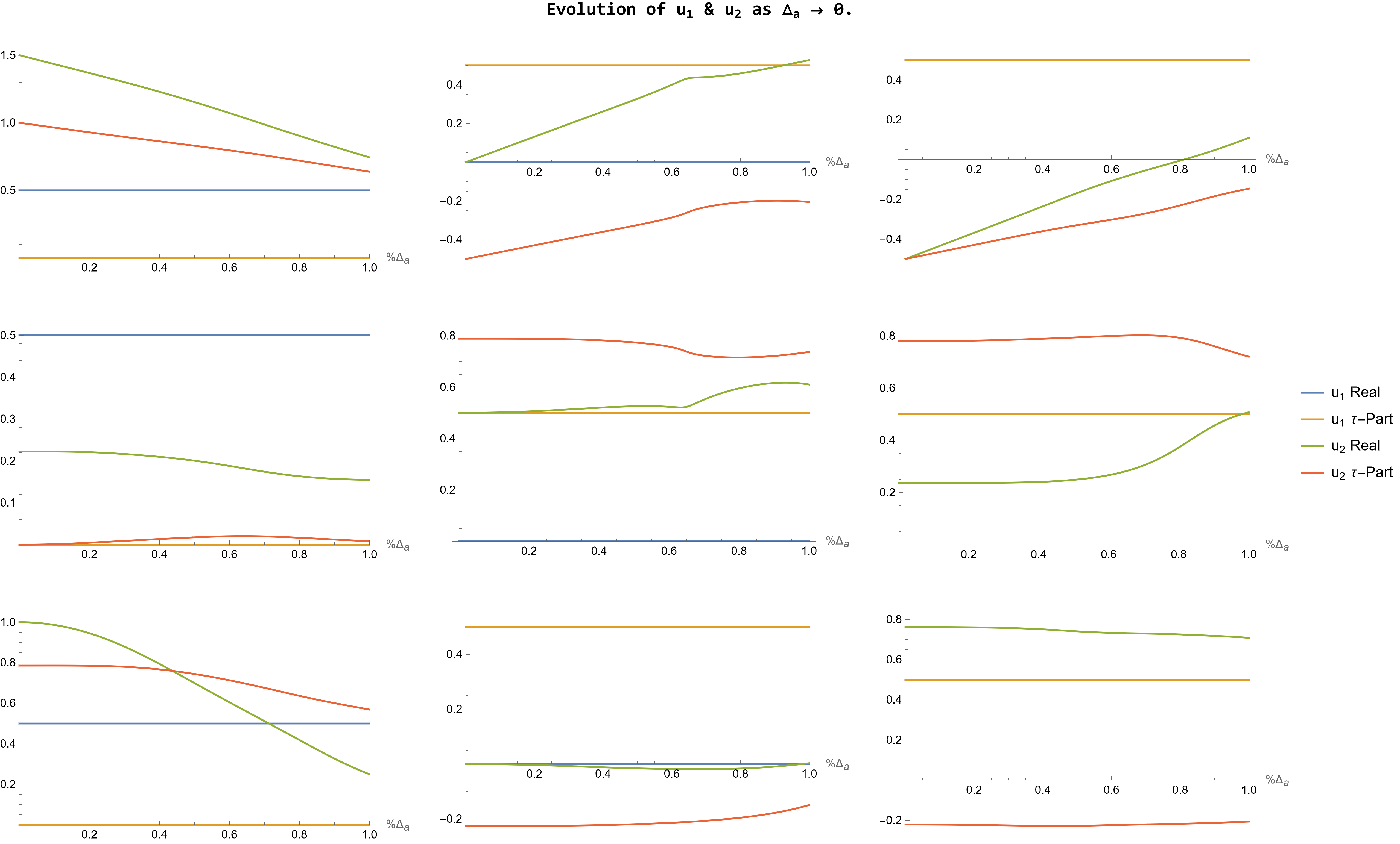}
  \caption{\justifying Tracking the nine half-rational $B_2$ BAE solutions as
  $\Delta_a\to0$ (the zero of the horizontal axis). The top row shows the three solutions flowing to
  poles of $V$; the lower two rows show the six flowing to the half-rational CM
  extrema. The starting values of $\Delta_a$ are chosen to be generic, $\Delta_1=0.421+0.411i$, $\Delta_2=0.34+0.33i$, while $\omega$ is held fixed at the generic value $\omega=1.82i$.}
  \label{fig:B2CMgrid}
\end{figure*}

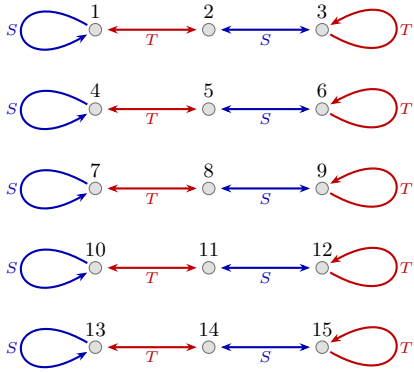
\begin{figure}[hbt!]
\centering
\begin{tikzpicture}[scale=0.75, every node/.style={scale=0.85}]
  \foreach \y [count=\row] in {0,-1.4,-2.8,-4.2,-5.6} {
    \begin{scope}[yshift=\y cm]
      \pgfmathtruncatemacro{\nOne}{(\row-1)*3+1}
      \pgfmathtruncatemacro{\nTwo}{(\row-1)*3+2}
      \pgfmathtruncatemacro{\nThree}{(\row-1)*3+3}
      \node[sol, label={[sollbl]90:$\nOne$}]   (N1) at (0,0) {};
      \node[sol, label={[sollbl]90:$\nTwo$}]   (N2) at (2.0,0) {};
      \node[sol, label={[sollbl]90:$\nThree$}] (N3) at (4.0,0) {};
      \draw[Sloop] (N1) to[out=180-30,in=180+30,looseness=40]
                   node[edgelbl, left] {$S$} (N1);
      \draw[Tedge] (N1) -- node[edgelbl, below]{$T$} (N2);
      \draw[Sedge] (N2) -- node[edgelbl, below]{$S$} (N3);
      \draw[Tloop] (N3) to[out=-30,in=30,looseness=40]
                   node[edgelbl, right]{$T$} (N3);
    \end{scope}
  }
\end{tikzpicture}
\caption{\justifying The $\mathrm{PSL}(2,\mathbb{Z})$ orbits of the $B_2$ BAE solutions. The five copies of the triplet are: solutions Nos. 1--3 are the fully-rational ones, solutions Nos. 13--15 are the half-pinned ones and solutions Nos. 4--12 are the half-rational ones. Finally, solutions Nos. 7--9 flow to poles of $V$, and all the other ones form the CM sector. Blue edges denote the $S$ generator, red edges $T$.}
\label{fig:B2CMPSL2Z}
\end{figure}

\paragraph{Half-rational solutions.}
There are nine further solutions, with a representative in which $u_1$ lies on
the half-lattice while $u_2$ is a function of $\omega$ and $\Delta_a$ obtained
as a root of a cubic. Given their complexity we display in figure \ref{fig:B2CMgrid} only the numerical
tracking of their flow as $\Delta_a\to0$, from the starting values $\omega=1.82i$,
$\Delta_1=0.421+0.411i$, $\Delta_2=0.34+0.33i$; the analytic check is
readily performed via computer algebra (\textsc{Mathematica}), taking due care of branch cuts.
Six of the nine flow to the six \emph{half-rational CM extrema}
\eqref{eq:B2halfrational}. The remaining three are our first examples of
nontrivial BAE solutions that map to poles of $V$. It is worth noting that
these three form a complete $\mathrm{PSL}(2,\mathbb{Z})$ orbit, see figure~\ref{fig:B2CMPSL2Z}.

\paragraph{\texorpdfstring{$C_2$}{C2}.}
Langlands duality relates $B_N$ and $C_N$ by exchanging long and short roots,
and at the level of groups exchanges the center with the fundamental group
\cite{Goddard:1976qe}, so that the simply connected form of one type
corresponds to the adjoint form of the other. For $B_2\cong C_2$ the two
algebras coincide, and the duality acts as a self-duality exchanging the two
global forms: the equations \eqref{eq:BAE_B2} are written in the orthogonal
basis spanning $\mathbb{Z}^2$, which is the adjoint form $\mathrm{SO}(5)$,
whereas the simply connected form $\mathrm{Spin}(5)\cong\mathrm{USp}(4)$
corresponds to the coroot lattice, of index two inside it. The $C_2$ solutions
and CM extrema therefore follow from those above by the corresponding change of
basis, which is precisely the Langlands-dual one.\footnote{In the string
realization of $\mathcal{N}=4$ SYM the two global forms arise from the
$\widetilde{\mathrm{O3}}^-$ and $\mathrm{O3}^+$ planes, related by the
$\mathrm{SL}(2,\mathbb{Z})$ action on the discrete torsions
\cite{Witten:1998xy,Hanany:2000fq}. We stress, however, that S-duality for
non-simply-laced algebras is not merely this exchange: it involves also the
Langlands map on the algebra and a rescaling of the holomorphic gauge coupling,
and is realized by a Hecke group rather than $\mathrm{SL}(2,\mathbb{Z})$
\cite{Argyres:2006qr}. The index does not depend on that coupling, so none of
this is visible here.} (See \cite{B2paper} for more details.)

\subsection{\texorpdfstring{The exceptional $G_2$ algebra}{The exceptional G2 algebra}}
\label{sub:G2}

For $G_2$ we have access only to numerical solutions of both the BAEs and the CM
extremum equations. We find 29 nontrivial BAE solutions and 15 CM extrema.
While we do not prove here that these lists are complete, a semi-analytic proof
will be presented in \cite{polypaper} using techniques from computational
commutative algebra. Accordingly we give here only the lists of solutions,
together with plots tracking them as $\Delta_a\to0$.

Of the 29 nontrivial BAE solutions, $14$ map to poles of $V$, while the
remaining $15$ constitute the CM sector, in bijection with the 15 CM extrema
found numerically.

%%%%%%%%%%%%%%%%%%%%%%%%%%%%%%%%%%%%%
\section{Conclusions}
\label{sec:conclusions}
%%%%%%%%%%%%%%%%%%%%%%%%%%%%%%%%%%%%%

We have exhibited a map from solutions of the $\mathcal{N}=4$ BAEs with
arbitrary semisimple gauge algebra $\alg$ to extrema of the untwisted elliptic
Calogero--Moser system of type $\alg$, obtained by following a solution
$u(\Delta_a)$ continuously as the chemical potentials $\Delta_a$ are sent to zero. The map
is well defined, it intertwines the symmetries of the two problems, and it
sends the trivial BAE solutions to poles of the CM potential. Its image,
however, is not all of the CM extrema: certain nontrivial solutions also flow
to poles, first at $B_2$, where three of the nine half-rational solutions do
so. Restricted to the remaining solutions, which we call the CM sector, we
conjecture the map to be a bijection, and we have verified this for every
rank-two semisimple algebra, classical and exceptional.

The system the map produces is untwisted for every $\alg$, and this is not a
choice on our part but a consequence of the structure of the BAEs, in which
every root enters with the same multiplicity and the same functional
dependence. For simply-laced algebras nothing is lost, since twisted and
untwisted coincide. For non-simply-laced algebras the twisted system is the one
governing the vacua of $\mathcal{N}=1^\ast$, and the map cannot reach it. The
conjectured bijection between $\mathcal{N}=4$ BAE solutions and
$\mathcal{N}=1^\ast$ vacua for type $A$ \cite{ArabiArdehali:2019orz,Benini:2021ano} therefore
does not extend beyond simply-laced type. What we propose in its place is the
same statement with the untwisted system substituted for the twisted one, a
replacement that is invisible in type $A$ and is forced everywhere else.

Several questions remain open. The most immediate is the interpretation of the
non-CM sector. We have no characterization of the solutions that flow to poles
beyond the observation that, for $B_2$, they form a complete
$\mathrm{PSL}(2,\mathbb{Z})$ orbit, so that the CM sector is itself a union of
orbits; whether this persists at higher rank, and whether these solutions are
distinguished by some property of the index rather than of the equations, we do
not know.

A second question is the extension to continuous families. The map applies to
them without modification, and we have checked for $A_2$ that the family of
\cite{Benini:2021ano} is exchanged with the family of CM extrema of
\cite{Bourget:2015upj} (see Appendix~\ref{sub:A2CM}). At rank two this is the only such case, so the evidence
is thin; since continuous families are expected to be generic at higher rank on
both sides, this is where the conjecture should next be tested. The polynomial
methods of \cite{polypaper}
make the higher-rank $\mathcal{N}=4$ BAE solutions accessible in
principle, and it would be worth computing the untwisted CM extrema for, say,
$A_3\cong D_3$ and $D_4$ to compare.

Finally, one may ask whether the twisted system has a Bethe-like counterpart of
its own. Our construction shows that the $\mathcal{N}=4$ BAEs relevant to compute the index do not produce
it. If the correspondence with $\mathcal{N}=1^\ast$ is to be restored outside
type $A$ (or any simply-laced type), some other set of equations must do so, and identifying them
would presumably require an object that distinguishes long from short roots in
a way the index does not. We leave this to future work.

\bigskip

\section*{Acknowledgments}

We are indebted to O.~Aharony for a careful reading of the draft; to A.~Amariti,
R.~Argurio, F.~Benini, A.~Bourget, R.~Klabbers, O.~Mamroud, S.~S.~Razamat,
A.~Torrielli and D.~Volin for useful discussions; to A.~Eghrari for
collaboration on a related project \cite{polypaper}; and to M.~De Marco and
A.~Zanetti for informing us of upcoming work on a related subject
\cite{demarco-zanetti}. This work has been supported by Royal Society
International Exchanges grant IES\textbackslash R2\textbackslash 242034, and
forms part of the Ph.D.\ project of K.K., supported by the Engineering and
Physical Sciences Research Council [Studentship 2934922]. We thank the
Universit\'e Libre de Bruxelles and the Universities of Milano and
Milano-Bicocca for hospitality during various stages of this work.

\appendix

%%%%%%%%%%%%%%%%%%%%%%%%%%%%%%%%%%%%%%%
\section{Derivations of the CM extrema}
\label{app:CM}
%%%%%%%%%%%%%%%%%%%%%%%%%%%%%%%%%%%%%%%

All potentials are written first in the orthogonal basis, and all solutions are
listed up to the Weyl action.

\subsection{\texorpdfstring{$A_2$}{A2}}
\label{sub:A2CM}

Twisted and untwisted CM systems coincide for simply-laced algebras, and the
$A_2$ extrema were already found in \cite[Eq. (4.16)]{Bourget:2015upj}. We quote the
isolated ones and slightly rewrite the continuous family. With the potential
\eqref{eq:A2pot}, the isolated extrema are the four points \eqref{eq:A2HL}.

The algebra $A_2$ is the only rank-two case admitting in addition a
one-dimensional family of extrema, at least among those we can treat
analytically; for $G_2$ our numerical results suggest none are present.
Writing $X_1\coloneqq\wp(q_1-q_2)$ and $X_2\coloneqq\wp(q_2-q_3)$, it is implicitly
given by \
\begin{equation}\label{eq:A2CMcont}
  X_1^2+X_1X_2+X_2^2=\tfrac{g_2}{4}\ .
\end{equation}

\subsubsection{The map on the continuous family}
\label{subsub:A2ContApp}

We start from \eqref{eq:A2BAEcont}. Under the map, $D\to1$, and setting
$z\coloneqq2Ku_{21}$, $w\coloneqq2Ku_{31}$, it moreover collapses to
\begin{equation}
  \cn(z)\,\cn(w)\,\cn(z-w)=1\ .
\end{equation}
Applying the standard addition formula to $\cn(z-w)$ gives
{\small
\begin{equation}
  \cn(z)\,\cn(w)\left(
  \frac{\cn(z)\cn(w)+\sn(z)\sn(w)\dn(z)\dn(w)}{1-mS_1S_2}\right)=1\ ,
\end{equation}}%
where $S_1\coloneqq\sn^2(z)$ and $S_2\coloneqq\sn^2(w)$. Clearing the
denominator, squaring and tidying up, we obtain
\begin{widetext}
\begin{equation}\label{eq:BAEcontAlmost}
  (1-mS_1S_2)\Big[S_1^2+S_1S_2+S_2^2-(m+1)S_1S_2(S_1+S_2)+mS_1^2S_2^2\Big]=0\ .
\end{equation}
\end{widetext}
Turning to \eqref{eq:A2CMcont}, we map the Weierstrass variables to Jacobi
functions via $X_i=e_3+(e_1-e_3)/S_i$, with the same $S_i$ and with
$e_1\coloneqq \wp(\tfrac12)$, $e_2\coloneqq \wp(\tfrac{\omega}{2})$,
$e_3\coloneqq \wp(\tfrac{1+\omega}{2})$ the roots of the Weierstrass cubic,
$\wp'^2=4(\wp-e_1)(\wp-e_2)(\wp-e_3)$, which satisfy $e_1+e_2+e_3=0$ and in
terms of which the elliptic parameter is $m=(e_2-e_3)/(e_1-e_3)$. Substituting,
simplifying with the standard Weierstrass identities and multiplying by
$S_1^2S_2^2$ to clear denominators yields
\begin{equation}\label{eq:CMCont}
  S_1^2+S_1S_2+S_2^2-(m+1)S_1S_2(S_1+S_2)+mS_1^2S_2^2=0\ .
\end{equation}
Equations \eqref{eq:BAEcontAlmost} and \eqref{eq:CMCont} agree up to the factor
$(1-mS_1S_2)$, whose zeros are the points at which the addition formula
degenerates: they enter only through the clearing of its denominator and are
not solutions of \eqref{eq:A2BAEcont}. The two continuous families are
therefore exchanged by the map.

\subsection{\texorpdfstring{$B_2\cong C_2$}{B2=C2}}
\label{sub:CMB2}

The untwisted $B_2$ CM potential is \eqref{eq:B2pot}. The computation of its
extrema below appears to be new, though it uses only standard techniques.

Using the Weierstrass addition formula, 
\begin{equation}
\wp(u+v)+\wp(u-v)=\dfrac{\wp'^2(u)+\wp'^2(v)}{2(\wp(u)-\wp(v))^2} -2\wp(u)-2\wp(v)\ ,
\end{equation}
and the curve equation $\wp'^2(z)=4\wp^3(z)-g_2\wp(z)-g_3$, we rewrite the
potential in terms of $x\coloneqq \wp(q_1)$ and $y\coloneqq \wp(q_2)$ as
\begin{equation}
  V(x,y)=\frac{4(x^3+y^3)-g_2(x+y)-2g_3}{2(x-y)^2}-x-y\ .
\end{equation}
The extremum equations become
$\partial_xV\,\wp'(q_1)=0$ and $\partial_yV\,\wp'(q_2)=0$, with
\begin{subequations}
{\small
\begin{align}
  E_1&\coloneqq\partial_{q_1}V=2x^3-6x^2y-6xy^2-6y^3+g_2x+3g_2y+4g_3\ ,\\
  E_2&\coloneqq\partial_{q_2}V=2y^3-6y^2x-6yx^2-6x^3+g_2y+3g_2x+4g_3\ .
\end{align}}
\end{subequations}
It is clear from this form that the extrema fall into three sectors.

\paragraph{Fully rational sector: \texorpdfstring{$\wp'(q_1)=\wp'(q_2)=0$}{}.}
Since $\wp'$ vanishes exactly at the three half-periods, $x$ and $y$ must be
roots of the Weierstrass cubic, conventionally $e_1,e_2,e_3$. As $x\neq y$ to
avoid the singularity at $q_1=q_2$, we obtain three distinct solutions,
\begin{equation}
  (x,y)\in\big\{(e_1,e_2),\,(e_2,e_3),\,(e_3,e_1)\big\}\ .
\end{equation}

\paragraph{Half-pinned sector: \texorpdfstring{$E_1=E_2=0$}{}.}
Adding the two equations gives a perfect cubic in $x+y$,
$(x+y)^3-g_2(x+y)-2g_3=0$, whose three roots are
\begin{equation}\label{eq:x+y}
  x+y=2e_k\ ,\qquad k\in\{1,2,3\}\ ,
\end{equation}
since the Weierstrass roots satisfy $(2e_k)^3-g_2(2e_k)-2g_3=0$. Subtracting
the two equations and dividing by $2(x-y)\neq0$ gives $x^2+xy+y^2=g_2/4$, from which we obtain
\begin{equation}\label{eq:xy}
  xy=(x+y)^2-(x^2+xy+y^2)=4e_k^2-\tfrac{g_2}{4}\ .
\end{equation}
Solving \eqref{eq:x+y} and \eqref{eq:xy} and eliminating $g_2$ via
$e_1e_2+e_2e_3+e_3e_1=-g_2/4$ and $e_1+e_2+e_3=0$ gives the three half-pinned
extrema
\begin{align}\label{eq:half-pinned}
  (x,y)=\Big(&e_k+i\sqrt{(e_i-e_k)(e_j-e_k)}\ ,\nonumber \\
  &e_k-i\sqrt{(e_i-e_k)(e_j-e_k)}\Big)\ .
\end{align}

\paragraph{Half-rational sector:
\texorpdfstring{$\wp'(q_1)=E_2=0$}{} (or vice versa).}
Setting $x=e_k$ and substituting into $E_2$,
\begin{equation}
  2y^3-6e_ky^2+(g_2-6e_k^2)y+10e_k^3-g_2e_k=0\ .
\end{equation}
The substitution $y=z+e_k$ removes the quadratic term, leaving
$z(2z^2+g_2-12e_k^2)=0$. The root $z=0$ gives $y=e_k$ and is discarded, so
\begin{equation}
  2z^2=12e_k^2-g_2=4\Big(3e_k^2-\tfrac{g_2}{4}\Big)=4(e_i-e_k)(e_j-e_k)\ .
\end{equation}
Each choice of $x=e_k$ thus admits two values of $y$, giving six distinct
half-rational extrema,
\begin{equation}\label{eq:half-rational}
  (x,y)=\Big(e_k,\ e_k\pm\sqrt{2(e_i-e_k)(e_j-e_k)}\Big)\ .
\end{equation}

\paragraph{Translation to Jacobi elliptic functions.}
To compare with the BAE solutions we rewrite the extrema in the original
variables $(q_1,q_2)$. For the fully rational sector the definitions
$e_1\coloneqq\wp(\tfrac12)$, $e_2\coloneqq\wp(\tfrac{\omega}{2})$,
$e_3\coloneqq\wp(\tfrac{1+\omega}{2})$ give, up to Weyl equivalence,
\begin{equation}\label{eq:B2CMfullyrational}
  (q_1,q_2) \in \left\{ \Big(\tfrac{1+\omega}{2},\tfrac{\omega}{2}\Big),\
  \Big(\tfrac{\omega}{2},\tfrac12\Big),\
  \Big(\tfrac{1+\omega}{2},\tfrac12\Big)\right\} \ .
\end{equation}
For the remaining sectors we need the standard relation
\begin{equation}
  \sn^2(2Kz,m)=\frac{e_1-e_3}{\wp(z)-e_3}\ ,
  \qquad m=\frac{e_2-e_3}{e_1-e_3}\ ,
\end{equation}
with $K$ the complete elliptic integral of the first kind. Evaluating at the
doubled argument $z=2q$ and isolating $q$ gives the inversion formula
\begin{equation}
  q=\varpi_a\pm\frac{1}{4K}\,\sn^{-1}
  \left(\sqrt{\frac{e_1-e_3}{\wp(2q)-e_3}},\,m\right),
\end{equation}
where $\varpi_a$ denotes any of the three half-periods, together with the
duplication formula
\begin{equation}
  \wp(2z)-e_3=\frac{\big((\wp(z)-e_3)^2-(e_1-e_3)(e_2-e_3)\big)^2}
  {4(\wp(z)-e_1)(\wp(z)-e_2)(\wp(z)-e_3)}\ .
\end{equation}

For the three half-pinned extrema \eqref{eq:half-pinned} the duplication
formula gives simply $\wp(2q)-e_3=-(e_1-e_3)(e_2-e_3)/(3e_3)$, from which we obtain the
convenient Weyl representatives, with $q_2=q_1+x$,
\begin{align}\label{eq:B2CMhalfpinned}
  q_1&=\tfrac{1}{4K}\sn^{-1}\big(\sqrt{2-m},m\big)\ , \quad\tfrac12\ ,\notag\\
  q_1&=\tfrac{1}{4K}\sn^{-1}\Big(\sqrt{\tfrac{1+m}{m}},m\Big)\ ,
  \quad x=\tfrac{\omega}{2}\ ,\\
  q_1&=\tfrac{1}{4K}\sn^{-1}\Big(\tfrac{\sqrt{2m-1}}{m},m\Big)\ ,
  \quad x=\tfrac{1+\omega}{2}\ .\notag
\end{align}
For the half-rational extrema \eqref{eq:half-rational} the manipulation is more
laborious but equally straightforward; the six independent extrema have
representatives
%\begin{widetext}
{\small
\begin{align}\label{eq:B2halfrational}
  &\left(\tfrac12,\ \varpi_a+\tfrac{1}{4K}\sn^{-1}
  \left(\sqrt{\tfrac{16-8m\pm12\sqrt{2(1-m)}}{17-9m\pm12\sqrt{2(1-m)}}},\,m
  \right)\right),\notag\\
  &\left(\tfrac{\omega}{2},\ \varpi_a+\tfrac{1}{4K}\sn^{-1}
  \left(\sqrt{\tfrac{16m-8\pm12i\sqrt{2m(1-m)}}
  {17m^2-9m\pm12im\sqrt{2m(1-m)}}},\,m\right)\right),\\
  &\left(\tfrac{1+\omega}{2},\ \varpi_a+\tfrac{1}{4K}\sn^{-1}
  \left(\sqrt{\tfrac{-8(m+1)\pm12\sqrt{2m}}{m}},\,m\right)\right).\notag
\end{align}}
%\end{widetext}

\subsection{\texorpdfstring{$D_2$}{D2}}
\label{sub:D2CM}

The untwisted $D_2$ potential is \eqref{eq:D2pot}. Since $D_2\cong A_1\oplus
A_1$, the change of basis $w_1\coloneqq q_1-q_2$, $w_2\coloneqq q_1+q_2$ brings it to a sum of
two $A_1$ potentials,
\begin{equation}
  V(w_1,w_2)=\wp(w_1)+\wp(w_2)\ .
\end{equation}
The extremum equations decouple into a pair of conditions $\wp'=0$, solved by
pairs of half-periods. Back in the original variables the nine extrema are:
\begin{widetext}
\begin{equation}\label{eq:D2sols}
  (q_1,q_2)\in\left\{\big(\tfrac12,0\big),\ \big(\tfrac{\omega}{2},0\big),\
  \big(\tfrac{1+\omega}{2},0\big),\
  \big(\tfrac{1+\omega}{4},\tfrac{3+\omega}{4}\big),\
  \big(\tfrac{1+\omega}{4},\tfrac{1+3\omega}{4}\big),\\
  \big(\tfrac{2+\omega}{4},\tfrac{\omega}{4}\big),\
  \big(\tfrac{2+\omega}{4},\tfrac{3\omega}{4}\big),\
  \big(\tfrac{1+2\omega}{4},\tfrac14\big),\
  \big(\tfrac{1+2\omega}{4},\tfrac34\big)\right\}\ .
\end{equation}
\end{widetext}

\subsection{\texorpdfstring{$G_2$}{G2}}
\label{sub:G2CM}

Unlike the classical cases treated in this Appendix, the $G_2$ extrema are known to us only numerically, and we do not reproduce their 15 sets of coordinates here; they are collected, together with the corresponding BAE solutions and the tracking data, in the \textsc{Mathematica} notebook attached to the arXiv submission. The $G_2$ holonomies in that notebook are expressed in the basis of fundamental coweights. 

\bibliography{baes}
\bibliographystyle{apsrev4-2}

\end{document}